\documentclass[prl,twocolumn,amsmath,amssymb,nofootinbib,preprintnumbers,balancelastpage]{revtex4}
\usepackage{graphicx}
\usepackage{hyperref}
\hypersetup{
    pdfnewwindow=true,     
    colorlinks=true,       
    linkcolor=black,       
    citecolor=blue,        
    filecolor=blue,        
    urlcolor=blue          
}
\begin{document}
\title{\Large{\bf{On the Origin of R-parity Violation in Supersymmetry}}
\footnote{Plenary talk given at 20th International Conference on Supersymmetry and Unification of Fundamental Interactions (SUSY2012), August 13-18, 2012, Beijing, China.}}

\author{\large{Pavel Fileviez P\'erez}}

\address{Particle and Astro-Particle Physics Division \\
Max Planck Institute for Nuclear Physics (MPIK) \\
Saupfercheckweg 1, 69117 Heidelberg, Germany}

%
\begin{abstract}
The fate of R-parity in supersymmetric theories is discussed in detail. We make a strong case for R-parity violation 
showing that the simplest theories based on the local B-L symmetry predict the spontaneous breaking of R-parity at the 
SUSY scale. The possible implications for the searches at the Large Hadron Collider are discussed.
\end{abstract}
\maketitle
%

\section{I. Introduction} 
The experimental collaborations at the Large Hadron Collider could discover Supersymmetry (SUSY) in the next years if this idea 
is realized at the TeV scale. Unfortunately, the simplest supersymmetric extension of the Standard Model (MSSM) 
cannot predict  the masses of the supersymmetric particles and very often the so-called R-parity discrete symmetry is assumed 
to be conserved. As is well-known the collider signatures of the MSSM depends of the violation or conservation of R-parity. 
Therefore, one can say that the fate of R-parity plays a major role in the discovery of SUSY at the LHC. 

In the context of the MSSM there are interactions violating the baryon or lepton numbers at the renormalizable level
\begin{equation}
{\cal W} \supset \epsilon \hat{L} \hat{H}_u \ + \ \lambda \hat{L} \hat{L} \hat{e}^c \ + \ \lambda^{'} \hat{Q} \hat{L} \hat{d}^c \ + \ \lambda^{''} \hat{u}^c \hat{d}^c \hat{d}^c.
\end{equation}
Here the first three interactions violate the total lepton number while the last term breaks the baryon number. 
When we combine the last two terms one gets the dimension four contributions to proton decay~\cite{review}. 
If these couplings, $\lambda^{'}$ and $\lambda^{''}$, are of order one and we assume low scale SUSY, 
where the mass of the squarks are at the TeV scale, we get a very small lifetime for the proton, $\tau_p \sim 10^{-20}$ years.
It is important to mention that the current experimental bounds on the proton decay lifetime are very strong, i.e. 
$\tau_p^{exp} > 10^{32-34}$ years. This critical issue has motivated different solutions: a) One can impose a discrete 
symmetry defined as $M=(-1)^{3(B-L)}$, which forbid {\textit{all}} the interactions above. $M$ is called matter-parity and is related 
to R-parity by the relation $M=(-1)^{2S} R$. In our notation, B and L are the baryon and lepton numbers, while S is the spin of the particle.
This solution is maybe too conservative because to avoid fast proton decay one does not need to forbid all the above couplings. 
b) One can assume that these couplings are very small, $ \lambda^{'} \lambda^{''} < 10^{-26}$, 
or c) Assume very large squark masses as in the case of split SUSY. Here we will discuss the possibility to understand dynamically 
the origin of these interactions and the possible predictions for the LHC searches. 

The simplest way to understand the origin of the B and L violating interactions in the MSSM is to consider a theory where 
$B-L$ is inside the algebra. The relation between $M$ parity and $B-L$ has been investigated by many groups~\cite{paper1,paper2,paper3,paper4,paper5,paper6,paper7,paper8}. It is important to say that most of the people in the SUSY community 
assume the conservation of matter parity because the lightest neutralino can describe the cold dark matter of the Universe. 
However, this is just one possibility and as I mentioned above, once you impose matter parity you just overkill the MSSM. 
\\
\\
The main aim of this talk is to prove that
\begin{center}
{\textit{The simplest theories based on local $B-L$ predict that R-parity {\bf{MUST}} be spontaneously broken !}}
\end{center}  
The main results presented here were found first in the context of a left-right symmetric theory~\cite{MinimalLR} and later 
investigated in the context of the minimal $B-L$ extension of the MSSM~\cite{MinimalB-L}.

\section{II. The Theory of R-parity} 

The simplest theory where one can understand the origin of the baryon and lepton number violating interactions in the MSSM at the renormalizable level 
is based on the gauge group~\cite{MinimalB-L} $$G_{B-L}=SU(3)_C \otimes SU(2)_L \otimes U(1)_Y \otimes U(1)_{B-L}.$$
The matter sector is composed of the MSSM chiral superfields
\begin{eqnarray}
&& \hat{Q} \sim (3,2,1/6,1/3), \  \hat{u}^c \sim (\bar{3},1,-2/3,-1/3), \nonumber \\ 
&& \hat{d}^c \sim (\bar{3},1,-1/3,-1/3), \  \hat{L} \sim (1,2,-1/2,-1), \nonumber \\
&&  \hat{e}^c \sim (1,1,-1,1), \  \rm{and} \ \hat{\nu}^c \sim (1,1,0,1). \nonumber
\end{eqnarray} 
Here, the right-handed neutrino superfields are needed to define an anomaly free theory. 
They cancel the anomalies $U(1)_{B-L}$ and $U(1)_{B-L}^3$. 

The Higgs superfields are the same as in the MSSM,
\begin{eqnarray}
\hat{H}_u \sim (1,2,1/2,0), \  {\rm{and}} \  \hat{H}_d \sim (1,2,-1/2,0). \nonumber
\end{eqnarray}
Therefore, one should worry about how to break the gauge symmetry $G_{B-L}$ to the SM symmetry.

The {\textit{only way}} to generate mass for the $B-L$ gauge boson in the minimal model is to generate 
a vacuum expectation value for the right-handed sneutrinos, i.e. $\left<\tilde{\nu}^c\right> \neq 0$. 
Then, one can say that the minimal $B-L$ model predicts the spontaneous breaking of R-parity 
and one should expect lepton number violation at the Large Hadron Collider if low energy 
supersymmetry is realized in nature.

The superpotential of this theory is quite simple
\begin{eqnarray}
{\cal {W}}_{B-L} &= &Y_u \hat{Q} \hat{H}_u \hat{u}^c +  Y_d \hat{Q} \hat{H}_d \hat{d}^c + Y_e \hat{L} \hat{H}_d \hat{e}^c \nonumber \\
&+&  Y_\nu \hat{L} \hat{H}_u \hat{\nu}^c +  \mu \hat{H}_u \hat{H}_d. 
\end{eqnarray}
Making use of the new $B-L$ contribution to the D-terms and the soft terms, one can prove that the vacuum expectation value 
(VEV) for the right-handed sneutrino is given by~\cite{MinimalB-L}
\begin{equation}
\left< \tilde{\nu}^c\right>=\sqrt{\frac{-4 M_{\tilde{\nu}^c}^2}{g_{BL}^2}}.
\end{equation}
Here $g_{BL}$ is the $B-L$ gauge coupling and $M_{\tilde{\nu}^c}$ is the soft mass for the $\tilde{\nu}^c$ field. 
Notice that in order to have a consistent symmetry breaking mechanism we need a negative $M_{\tilde{\nu}^c}^2$.  
The same is assumed for the Higgs in the Standard Model. As one expects the R-parity and $B-L$ breaking scales are determined by the SUSY breaking scale.

After symmetry breaking one predicts the existence of lepton number violating couplings at the renormalizable level, 
while the baryon number violating terms are generated at the non-renormalizable level~\cite{RpVLHC}. Therefore, assuming a large cutoff or the 
great desert, one can satisfy the bounds on the proton decay lifetime. The same has to be assumed for the allowed dimension 
five operators $\hat{Q} \hat{Q} \hat{Q} \hat{L} / \Lambda$. See Ref.~\cite{review} for the study of these operators.

There are two simple ways to explain dynamically the negative mass squared of the right-handed sneutrinos. 
Assuming that the soft masses for the right-handed sneutrinos of each generation are 
different at the grand unified scale, it is possible to generate a negative mass squared for one of them at the low scale. 
For details see Refs.~\cite{Ovrut1,Ovrut2,Ovrut3}. The second possibility is to go beyond the minimal model and 
assume that the right-handed neutrinos couple to a new Higgs with a large coupling~\cite{Fate1,Fate2}.
\\
\\
{\textit{How do we test the predictions coming from this theory? }}

As in any gauge extension of the SM one has new gauge bosons. 
In this case the mass of the new $B-L$ gauge boson is predicted by the SUSY scale. 
At the same time since R-parity and lepton number is broken after symmetry breaking one 
should expect some exotic signals at the LHC. We have investigated this issue 
in Ref.~\cite{RpVLHC} and proposed the channels with multileptons as possible signals coming from this theory.
In the case when the lightest neutralino is the lightest supersymmetric particle one can think about the pair slepton production 
through the photon, the $Z$ and the $B-L$ gauge boson $Z_{B-L}$,
\begin{eqnarray}
p \ p \  &\to&  \  \gamma, Z, Z_{B-L} \  \to \ \tilde{e}_i^* \tilde{e}_i \  \to e_i^+ e_i^- \tilde{\chi}_1^0 \tilde{\chi}_1^0 \nonumber \\
&\to & \   e_i^+ e_i^- e^{\pm}_j e^{\pm}_k W^{\mp} W^{\mp} \to  e_i^+ e_i^- e^{\pm}_j e^{\pm}_k 4 j. \nonumber
\end{eqnarray}
Therefore, one can have channels with four leptons, three of them with the same electric charge, and in order to look for lepton number violation one should use the jets from the W decays. The numerical predictions for the cross section and branching ratios have been investigated in Ref.~\cite{RpVLHC}. Since the background for these exotic signatures is suppressed one can hope to discover or set strong bounds on the masses of these particles. 
It is important to mention that in this context the gravitino can be the dark matter candidate or one 
can use the idea proposed in Ref.~\cite{Jon}, where the dark matter sector is supersymmetric. 

We have mentioned that the main idea discussed here was pointed out for the first time in the context of left-right symmetric 
theories~\cite{LR1,LR2,LR3,LR4}. In Ref.~\cite{MinimalLR} we have shown that in the minimal supersymmetric left-right 
symmetric theory based on $$SU(2)_L \otimes SU(2)_R \otimes U(1)_{B-L},$$ one must have the spontaneous breaking of R-parity. In this context the leptonic superfields are given by
\begin{align}
 \hat{L} &= \begin{pmatrix}
                     \hat{\nu}  \\
                     \hat{e} \\
                    \end{pmatrix} \sim (2,1,-1),   
                    \text{ and } 
 \hat{L}^c = \begin{pmatrix}
                     \hat{\nu}^c \\
                     \hat{e}^c  \\
                    \end{pmatrix} \sim (1,2,1).
\end{align}
Since the Higgs present in the minimal model is the bidoublet, $\hat{\Phi} \sim (2,2,0)$, the only way to break the $SU(2)_R \otimes U(1)_{B-L}$ 
symmetry is to give a vacuum expectation value to the right-handed leptonic double $\hat{L}^c$. Here the VEV of the right-handed sneutrino is given by
\begin{equation}
v_R = \sqrt{\frac{-8 M_{\tilde{L}^c}^2}{g_{BL}^2 + g_R^2}}.
\end{equation}
The coupling $g_R$ is the gauge coupling associated to $SU(2)_R$ and $M_{\tilde{L}^c}$ is the soft mass for the leptonic doublet $\tilde{L}^c$. We should mention that 
in order to achieve the symmetry breaking  in this context, we need to assume that the soft terms in the leptonic sector break the left-right discrete symmetry.
As one can appreciate here,  we can get a similar result as in the minimal $B-L$ model discussed above. If we stick to the minimal supersymmetric left-right 
model R-parity must be spontaneously broken. 
\\
\\
{\textit{What about R-parity conservation?}} 

If the minimal $B-L$ theory predicts R-parity violation, we need to go beyond the simple model to explain dynamically the conservation of R-parity. 
Therefore, we need to add extra Higgses with even number of $B-L$~\cite{paper6} or use the Stueckelberg mechanism~\cite{Feldman} to generate mass for the new gauge boson. In both cases one needs 
to do more model building. 

This issue has been discussed in the context of left-right models. In order to achieve the conservation of R-parity you need to add many Higgses, 
see Refs.~\cite{LRRpC1,LRRpC2}. In the context of $SU(5)$ we cannot explain the conservation of R-parity, while in $SO(10)$ you need to assume a very involved Higgs sector~\cite{paper8}. 
Since these theories are not very appealing, I do not understand why the people in the SUSY community always assume R-parity conservation. This scenario can be simple 
but not very well-motivated from the theoretical point of view.  

Finally, I would like to mention a more radical scenario where there is no need to assume the existence of a great desert to suppress the dimension five operators 
mediating nucleon decay. One can define a simple theory where B and L are local symmetries that are spontaneously broken at the low scale. We have shown in 
Refs.~\cite{Vectorlike,Duerr} the different possibilities. The details of this discussion are beyond the scope of this talk.  

\section{III. Summary}

In this talk I have discussed that the minimal theories based on the $B-L$ gauge group predict that R-parity must be spontaneously broken and one expects lepton number violating signatures at the LHC.
It is important to emphasize that there are no simple grand unified theories defined in four dimensions which predict the conservation of R-parity in Supersymmetry. 
Therefore, one can say that there is a lack of theoretical motivations for R-parity conservation. In summary, I would like to repeat that the spontaneous breaking of R-parity is well motivated 
and could be crucial for the discovery of Supersymmetry at the LHC. I hope that these results will motivate more experimental studies for R-parity violating signatures at the Large Hadron Collider. 
\\
\\
{\textit{Acknowledgments}}:
I thank the organizers of SUSY2012 for the invitation, the warm hospitality and an excellent conference. I am grateful to my collaborators 
Vernon Barger, Pran Nath, Sogee Spinner, and Mark B. Wise for many discussions and great collaborations.

\end{document}